# Stato evolutivo delle stelle della Cintura di Orione ed implicazioni archeoastronomiche

Vincenzo Orofino
(Dipartimento di Fisica, Università del Salento, Lecce, Italy)

**Abstract**
The Orion Belt is a three-star linear asterism located in the central part of the Orion constellation which was associated by the ancient Egyptians to god Osiris.
The Orion Belt is the subject of a controversial theory proposed by Bauval e Gilbert (1994), the so-called "Orion Correlation Theory" (OCT). According to these authors, a perfect coincidence would exist between the disposition of the three stars of the Orion Belt and that of the main Giza pyramids, so that the latter would represent the monumental reproduction on the ground of that important asterism (see Bauval, 2006).
In a companion paper (Orofino, 2011) the OCT has been reanalyzed, subjecting it to some quantitative astronomical and astrophysical verifications, in order to assess the compatibility of this theory with the results of both naked-eye astrometry and photometry. In that work it has been found that the relative positions of the three Giza pyramids coincide, within the uncertainties of the naked-eye astrometric measurements, with the relative positions of the three stars of the Orion Belt. Furthermore, a simple and interesting correlation has been found between the visual magnitude of the stars of the Belt and the height of the corresponding pyramids evaluated with respect to a common reference level (i.e. the sea level or the base level of the Khufu pyramid).
Now, another question remains to be evaluated: whether at the time of the pyramids the magnitudes of the Orion Belt stars were different from the present ones, implying that the correlation found by Orofino (2011) could not be valid at that epoch. Actually, this hypothesis is not unfounded at all; in fact Mintaka, Alnitak and Alnilam (especially the last two) are blue supergiants that could be in a very advanced evolutionary phase, characterized by a nucleus with an hydrogen content close to the exhaustion or just exhausted. When the nuclear fuel of a star begins to run low, the consequent reduction of the energy flux coming from the nucleus causes irregular variations of the radius and of the superficial temperature; for this reason the star shows relatively conspicuous and fast changes in its luminosity. Such variations in luminosity, instead, would not be possible in the case of relatively unevolved objects (such as our Sun), which are very common in our Galaxy.
In the present work the evolutionary status of the Orion Belt stars is evaluated. It is found that, according to both kinds of stellar evolution models – the traditional ones (Lejeune e Schaerer, 2001), that do not take into consideration the star rotation, and the more recent ones (Meynet e Maeder, 2005) that take it into consideration – the three stars of the Orion Belt are evolved but stationary objects, in the sense that their visual magnitude has remained practically the same, at least in the last 10 thousand years; for this reason the correlation found by Orofino (2011) was still valid when the pyramids were built.
It is important to note that the constancy of the magnitude of the Orion Belt stars in the last 10 thousand years is not obvious at all, as one could expect taking into consideration the evolution of solar-type stars. Actually, if Alnilam were only 8000 years older, then, according to the traditional models, its present magnitude would be substantially different from that at the time of pyramids, and the correlation found by Orofino (2011) would be no more valid at that time.
----------
Nel presente lavoro viene determinato lo stato evolutivo delle stelle della Cintura di Orione, ricavando che, all'epoca della costruzione delle piramidi, la luminosità delle tre stelle della Cintura era di fatto uguale a quella odierna. Tale non banale risultato riveste una importanza fondamentale nell'ambito della verifica della controversa Teoria della Correlazione di Orione proposta da Bauval e Gilbert nel 1994, secondo la quale esisterebbe una perfetta coincidenza tra la disposizione delle tre stelle della Cintura e quella delle tre piramidi nella piana di Giza.

## 1. Introduzione

La Cintura di Orione è un asterismo lineare posto all'interno dell'omonima costellazione e costituito da tre stelle delle quali quella centrale è grosso modo equidistanziata dalle due estreme, ma con la stella più settentrionale (Mintaka) leggermente fuori asse di circa 4° rispetto alla congiungente le altre due (Alnitak, la più meridionale, ed Alnilam, l'astro centrale).

Anche se non mancano interpretazioni alternative (Baux, 1993; Legon, 1995), è comunemente ritenuto che gli antichi Egizi associassero la Cintura di Orione, e più in generale la costellazione di Orione, al dio Osiride, una loro importantissima divinità (si veda Bauval, 2006 ed i lavori ivi citati). Secondo alcuni autori (Gribbin e Gribbin, 1996; Tedder, 2006) tale asterismo è stato rappresentato nell'affresco sul soffitto famosa tomba di Senmut, un dignitario della regina Hatshepsut (ca. XV secolo a.C.). In questo dipinto Orione (Sah) è raffigurato su una barca, immediatamente al di sotto di un gruppo quasi verticale di tre stelle che ricorda molto da vicino la Cintura (la concordanza è, infatti, rimarchevole a parte il dettaglio, irrilevante per un disegno che non è una mappa, di un angolo di disallineamento di 7°, invece dei 4° della realtà).

La Cintura di Orione è oggetto di una suggestiva ma estremamente controversa ipotesi, nota con il nome di Teoria della Correlazione di Orione (TCO), proposta da Bauval e Gilbert (1994). Secondo tali autori esisterebbe una perfetta coincidenza tra la disposizione delle tre stelle della Cintura e quella delle tre piramidi nella piana di Giza, e pertanto quest'ultime rappresenterebbero la monumentale riproduzione al suolo di quell'importante asterismo (si veda anche Bauval, 2006).

In un recente articolo (Orofino, 2011) quest'ipotesi è stata ripresa in considerazione, sottoponendo l'intera TCO ad alcune verifiche di natura astronomica, allo scopo di verificarne la compatibilità con quanto previsto sia dall'astrometria che dalla fotometria stellare ad occhio nudo. In tale lavoro è stato innanzitutto provato che le posizioni relative delle tre piramidi di Giza coincidono, entro le incertezze delle misure astrometriche del tempo, con le posizioni relative dei tre astri della Cintura di Orione; è stato inoltre provato che nella scala che sarebbe stata scelta dai costruttori delle piramidi per riprodurre al suolo l'asterismo della Cintura, la distanza lineare che separa la piramide di Micerino dal centro del Nilo lungo la congiungente le piramidi estreme coincide con la distanza angolare che separa Mintaka dal punto centrale della Via Lattea lungo la congiungente le due stelle estreme dell'asterismo. E' stata infine ricavata una semplice quanto interessante correlazione che lega la magnitudine delle stelle della Cintura con l'altezza delle corrispondenti piramidi rispetto ad un comune livello di riferimento (p.e. il livello di base della piramide di Cheope).

Un ulteriore test cui deve essere sottoposta la sopra discussa correlazione consiste nel verificare se per caso in passato le magnitudini delle stelle della Cintura fossero diverse da quelle attuali, implicando che tale correlazione potesse non essere rispettata a quel tempo. Questa ipotesi non è, in verità, priva di fondamento. Effettivamente Mintaka, e soprattutto Alnitak ed Alnilam, sono stelle supergiganti blu che potrebbero trovarsi in uno stadio evolutivo molto avanzato, caratterizzato da un contenuto di idrogeno nel loro nucleo in fase di esaurimento oppure appena esaurito. Quando il combustibile nucleare di una stella inizia a scarseggiare, la conseguente riduzione del flusso di energia proveniente dal nucleo provoca delle irregolari variazioni del raggio e della temperatura superficiale; di conseguenza la stella subisce relativamente cospicui e rapidi cambiamenti di luminosità. Ciò potrebbe causare un odierno valore della magnitudine di questi oggetti diverso da quello al tempo degli antichi Egizi. Simili variazioni di luminosità non sarebbero invece possibili nel caso di sorgenti poco evolute, il che è una situazione estremamente comune tra le stelle della Galassia.

Scopo del presente lavoro è appunto quello di effettuare la suddetta verifica determinando lo stato evolutivo delle tre stelle della Cintura.

## 2. Modelli evolutivi

L'evoluzione di una stella dipende in modo cruciale dalla sua composizione, in particolare dipende dalla cosiddetta *metallicità* Z dell'oggetto, ossia dall'abbondanza percentuale (relativa all'idrogeno) degli elementi più pesanti dell'elio, quali litio, berillio, boro, ecc. In letteratura esistono tracciati evolutivi teorici per stelle di varia metallicità tra Z = 0.001 e Z = 0.1. In generale per le stelle della nostra Galassia tale abbondanza relativa è compresa tra 0.002 e 0.1 (Nordström et al., 2004), valendo Z = 0.02 nel caso del Sole; in particolare un campione di giganti e supergiganti blu analizzato da Niemczura (2003) presenta valori di Z tra 0.005 e 0.03 con un valor medio di 0.01. Sfortunatamente la metallicità delle tre stelle della Cintura non è al momento nota.

I primi modelli d'evoluzione stellare sono stati sviluppati negli anni '90 in una serie di lavori discussi e riassunti da Lejeune e Schaerer (2001). Questi modelli descrivono oggetti con velocità di rotazione assiale trascurabile ed hanno un output dettagliato presentato sotto forma di tabelle consultabili on-line al sito http://vizier.cfa.harvard.edu/viz-bin/VizieR?-source=VI/102. Tali tabelle forniscono l'evoluzione temporale di diversi parametri stellari (tra cui tutte le grandezze riportate nella Tabella 1) per stelle di metallicità compresa tra Z = 0.0004 e Z = 0.1 e di massa iniziale compresa tra 0.8 e 120 $M_\odot$.

Oltre a questi modelli senza rotazione, a partire dai primi anni del 2000 sono stati sviluppati anche modelli con stella rotante (si veda Meynet e Maeder (2005) e referenze ivi incluse); questi ultimi sono più completi di quelli tradizionali ma, almeno per quanto riguarda i risultati pubblicati, hanno un output meno dettagliato dei precedenti. Quello che si può fare con questi modelli è confrontare i valori teorici di $x = \log(T_e)$ ed $y = \log(L/L_\odot)$ con i corrispondenti dati osservativi $x_0$ ed $y_0$, collocando questi ultimi in un grafico *x-y* (diagramma H-R) dove sono rappresentati i tracciati evolutivi previsti dai modelli per stelle di una data massa iniziale e di una data metallicità. In particolare i dati osservativi possono essere riportati nei grafici della Figura 3 di Meynet e Maeder (2005) dove sono appunto mostrati i tracciati evolutivi di due stelle di massa iniziale pari 40 $M_\odot$ e 120 $M_\odot$ ottenuti per vari valori di Z tramite un modello che prevede per entrambe una velocità di rotazione iniziale di 300 km/s (all'equatore). Si può così determinare se i dati osservativi sono consistenti con quelli attesi per una stella con un dato valore di *M* e di Z.

## 3. Attuale fase evolutiva delle stelle della Cintura

Nella Tabella 1 sono riportati i valori dei principali parametri stellari relativi ad Alnitak, Alnilam e Mintaka. Tra le grandezze ivi riportate compaiono, tra l'altro, la massa iniziale dell'oggetto (cioè quella al momento dell'innesco delle reazioni di fusione nucleare dell'idrogeno nelle zone centrali della stella) e la magnitudine visuale assoluta, ossia la magnitudine che l'oggetto avrebbe se si trovasse ad una distanza di 10 pc dalla Terra. Nel caso di Alnitak ed Alnilam i dati stellari provengono principalmente da studi dettagliati sulle stelle in questione (Bouret et al. (2008) per Alnitak; Blomme et al. (2002) per Alnilam), mentre nel caso di Mintaka, mancando studi dettagliati, i dati provengono dal lavoro di Lamers e Leitherer (1993) che hanno studiato un campione di 28 stelle comprendente le tre stelle della Cintura. Per tutti e tre gli oggetti i dati stellari (con i loro errori) sono stati completati e/o aggiornati nel presente lavoro.

**Tabella 1** – Dati stellari di riferimento per le tre stelle della Cintura di Orione. Le varie colonne riportano rispettivamente il nome dell'oggetto, il logaritmo della temperatura efficace (in kelvin), il logaritmo della luminosità (in unità della luminosità solare $L_\odot$), il logaritmo della gravità superficiale (in cm s$^{-2}$), la magnitudine visuale assoluta, la massa iniziale (in unità della massa solare $M_\odot$) ed infine il logaritmo del tasso di perdita di massa (in masse solari all'anno).

| Stella | log ($T_e$) | log ($L/L_\odot$) | log ($g$) | $M_V$ | $M$ | log ($\dot{M}$) |
|---|---|---|---|---|---|---|
| Alnitak | $4.470^{+0.014}_{-0.015}$ | $5.64 \pm 0.15$ | $3.25 \pm 0.10$ | $-6.35 \pm 0.37$ | $(40 \pm 20)\, M_\odot$ | $-5.78 \pm 0.15$ |
| Alnilam | $4.455^{+0.015}_{-0.016}$ | $5.58 \pm 0.20$ | $3.06 \pm 0.10$ | $-6.62 \pm 0.50$ | $35^{+25}_{-8}\, M_\odot$ | $-5.73 \pm 0.15$ |
| Mintaka | $4.517^{+0.030}_{-0.030}$ | $5.82 \pm 0.20$ | $3.12 \pm 0.20$ | $-6.57 \pm 0.50$ | $45.0^{+11.2}_{-9.5}\, M_\odot$ | $-5.54 \pm 0.23$ |

Una volta individuati i valori dei parametri stellari dei tre oggetti della Cintura, così come dedotti dalle osservazioni, si è cercato di determinare, utilizzando i modelli d'evoluzione stellare, a quale stato evolutivo corrispondessero valori teorici dei parametri stellari in grado di riprodurre meglio quelli dedotti dalle osservazioni (d'ora in poi chiamati, anche se impropriamente, osservativi).

### 3.1 Alnitak (ξ Orionis)

Nel caso di Alnitak, sfruttando le tabelle on-line relative ai modelli senza rotazione stellare si ricava che i dati osservativi riportati nella Tabella 1 vengono ottimamente riprodotti da un oggetto di età pari a 3.9 milioni di anni, di metallicità Z = 0.02 e di massa attualmente pari a 37.4 $M_\odot$ ancora lontano dall'esaurimento dell'idrogeno nel nucleo e che pertanto si trova in uno stato di luminosità praticamente costante. Si noti che in tal caso la metallicità di Alnitak risulta del tutto consistente con quella del campione di giganti blu analizzato da Niemczura (2003). Al contrario non esistono oggetti evoluti, che si trovano in uno stato di luminosità apprezzabilmente variabile su tempi-scala dell'ordine dei 5000 anni (ossia l'intervallo di tempo che ci separa dalla costruzione delle piramidi), i cui parametri stellari sono in grado di riprodurre in modo accettabile i dati osservativi.

Sempre nel caso di Alnitak, anche i modelli con stella rotante forniscono risultati simili a quelli ottenuti con i modelli tradizionali. Infatti, se sui suddetti tracciati evolutivi ottenuti da Meynet e Maeder (2005) si riportano i dati osservativi $x_0$ ed $y_0$ relativi a questa stella (v. Tabella 1), si vede che, indipendentemente dalla metallicità, questi ultimi sono meglio riprodotti da un oggetto di massa iniziale pari a 40 $M_\odot$ posto in uno stato stazionario (ossia dotato di temperatura e luminosità costanti su tempi-scala dell'ordine almeno dei 10 mila anni).

**3.2 Alnilam (ε Orionis)**

Nel caso di Alnilam, sfruttando le tabelle on-line riassuntive dei risultati dei modelli senza rotazione (Lejeune e Schaerer, 2001), si ricava che i dati osservativi riportati nella Tabella 1 vengono meglio riprodotti da un oggetto di 4.8 milioni di anni, di metallicità Z = 0.008 (valore del tutto consistente con quella del campione di giganti blu analizzato da Niemczura, 2003) e di massa attualmente pari a 37.8 $M_\odot$ che è molto prossimo all'esaurimento dell'idrogeno nucleare ma che si trova ancora in uno stato di luminosità praticamente costante. Anche in questo caso non esistono oggetti evoluti, che si trovano in uno stato di luminosità apprezzabilmente variabile su tempi-scala dell'ordine dei 5000 anni, i cui parametri stellari sono in grado di riprodurre in modo accettabile i dati osservativi.

Per quanto riguarda invece i modelli con stella rotante, si vede che il tracciato evolutivo che più si avvicina al punto osservativo ($x_0$, $y_0$) è quello relativo ad una stella di 40 $M_\odot$ e Z = 0.02 che si trova in uno stato stazionario. In altre parole, anche per Alnilam con entrambi i tipi di modelli gli stati stazionari riproducono i dati osservativi molto meglio di quelli non stazionari; questi ultimi sono quindi alquanto improbabili

**3.3 Mintaka (δ Orionis)**

Nel caso di Mintaka le tabelle on-line relative ai modelli con stella non rotante (Lejeune e Schaerer, 2001) possono essere utilizzate interpolando fase per fase i dati relativi a stelle di massa iniziale pari a 40 $M_\odot$ e a 60 $M_\odot$ allo scopo di ottenere quelli relativi ad una stella di 50 $M_\odot$. Si ricava così che i dati osservativi riportati nella Tabella 1 sono meglio riprodotti da quelli di un oggetto di 3.9 milioni di anni, di metallicità Z = 0.008 e di massa attualmente pari a 47.4 $M_\odot$ che si trova in uno stato di luminosità praticamente costante, in quanto è ancora lontano dall'esaurimento dell'idrogeno nel proprio nucleo.

Come nei casi precedenti, per determinare lo stato evolutivo di Mintaka si possono utilizzare anche i modelli con stella rotante (Meynet e Maeder, 2005), seguendo lo stesso procedimento già seguito per Alnitak ed Alnilam. Prima però è necessario eseguire una doppia interpolazione dei dati teorici, sia su Z che su *M*. Seguendo questa procedura si ricava che, quando un oggetto di massa iniziale 50 $M_\odot$ e Z = 0.008 si trova in una fase stazionaria caratterizzata da una temperatura pari a quella osservata per Mintaka, allora la sua luminosità è tale che $\log(L/L_\odot) = 5.79$, in ottimo accordo con le osservazioni che danno $\log(L/L_\odot) = 5.82 \pm 0.20$. In conclusione, sia i modelli stellari non rotanti che quelli rotanti suggeriscono che anche Mintaka sia un oggetto in uno stato stazionario e che quindi abbia una temperatura ed una luminosità costanti su tempi-scala dell'ordine di almeno 10 mila anni.

**4. Discussione e conclusioni**

Riassumendo le analisi sopra riportate per le tre stelle della Cintura, si può concludere che i valori dei parametri riportati nella Tabella 1 sono meglio riprodotti (come indicato chiaramente dal confronto tra i $\chi^2$ dei fit dei dati osservativi ottenuti con un oggetto stazionario e con uno non stazionario) da oggetti che, per quanto evoluti, si trovano ancora abbastanza lontani dall'esaurimento dell'idrogeno centrale; e questa conclusione vale indipendentemente dal fatto di utilizzare modelli stellari con oppure senza rotazione. In altre parole, secondo quanto suggerito da entrambi i tipi di modelli stellari, tutte e tre sono oggetti stazionari e pertanto la loro magnitudine visuale deve essere rimasta di fatto costante almeno negli ultimi 10 mila anni.

In ogni caso Alnilam sembra essere la più evoluta tra le tre stelle. In effetti i best fit dei dati osservativi di Mintaka ed Alnitak ottenuti utilizzando i modelli tradizionali sono entrambi relativi ad oggetti che si trovano in una fase evolutiva compresa tra gli stati 8 e 9 delle tabelle on-line sopra discusse (Lejeune e Schaerer, 2001); il best fit dei dati osservativi di Alnilam è relativo invece ad un oggetto posto tra gli stati 11 e 12, cioè che si trova ancora in uno stato stazionario, ma che è molto prossimo all'esaurimento dell'idrogeno nucleare. La stella dovrebbe infatti restare in questa fase relativamente stabile soltanto per 4000 anni, superati i quali essa dovrebbe subire una rapida variazione di magnitudine (circa pari a 0.14 in soli 4000 anni).

D'altra parte la sovrapposizione dei valori osservativi di $\log(T_e)$ e di $\log(L/L_\odot)$ sui tracciati evolutivi ottenuti utilizzando i modelli con stella rotante, indica che lo stato evolutivo di Alnilam è più vicino, rispetto a quelli di Mintaka ed Alnitak, al punto del diagramma H-R, corrispondente al massimo relativo di $T_e$, a sua volta abbastanza prossimo al punto di arresto della combustione nucleare dell'idrogeno centrale.

Quindi Alnilam dovrebbe essere un oggetto sicuramente più evoluto delle altre due stelle della Cintura, anche se i due tipi di modelli non forniscono le stesse indicazioni circa l'effettiva fase evolutiva della stella: prossima all'esaurimento dell'idrogeno nucleare nel caso dei modelli tradizionali, un po' più lontana da questo stadio, nel caso dei modelli con stella rotante. Tali modelli sono comunque d'accordo sul fatto che Alnilam si deve attualmente trovare ancora in uno stato stazionario.

E' importante sottolineare il fatto che l'ipotesi sottoposta ad esame in questo lavoro, ossia che la magnitudine delle tre stelle della Cintura di Orione si sia mantenuta costante nei 5000 anni che ci separano dalla costruzione delle piramidi, non è affatto ovvia come si potrebbe pensare prendendo ad esempio l'evoluzione di stelle come il Sole. In effetti, in base ai modelli tradizionali, se Alnilam avesse soltanto 8000 anni in più rispetto alla sua età reale (un'inezia su scala astronomica), allora la sua magnitudine attuale sarebbe sostanzialmente diversa da quella all'epoca delle piramidi e la correlazione trovata da Orofino (2011) a quel tempo non sarebbe stata verificata.